\documentclass[prd,10pt,aps,showpacs,superscriptaddress,twocolumn,amsmath,amssymb,nofootinbib,longbibliography]{revtex4-1}

\usepackage{natbib}
\usepackage[colorlinks]{hyperref}
\hypersetup{linkcolor=blue,citecolor=blue}

\usepackage{graphicx}
\usepackage{dcolumn}
\usepackage{bm}
\usepackage{lipsum} 
\usepackage{color}
\usepackage{soul}

\newcommand{\mnras}{MNRAS}

\def\apj{ApJ}
\def\mnras{MNRAS}

\def\prd{PRD}

\def\nat{Nature}

\begin{document}

\title{Exploring light propagation in nonlinear electrodynamics: phase and group velocities and related phenomena}

\author{T. W. \surname{Cruz}}
\email{thwcruz@ifi.unicamp.br}
\affiliation{Instituto de F\'{\i}sica e Qu\'{\i}mica, Universidade Federal de Itajub\'a, \\
Itajub\'a, Minas Gerais 37500-903, Brazil}
\affiliation{Instituto de Física Gleb Wataghin, Universidade Estadual de Campinas,\\ Campinas, São Paulo 13083-859,
Brazil}
\author{V. A. \surname{De Lorenci}}
\email{delorenci@unifei.edu.br}
\affiliation{Instituto de F\'{\i}sica e Qu\'{\i}mica, Universidade Federal de Itajub\'a, \\
Itajub\'a, Minas Gerais 37500-903, Brazil}
\author{E. \surname{Guzm\'an-Herrera}}
\email{elda.guzman@unifei.edu.br}
\affiliation{Instituto de F\'{\i}sica e Qu\'{\i}mica, Universidade Federal de Itajub\'a, \\
Itajub\'a, Minas Gerais 37500-903, Brazil}
\author{C. C. H. \surname{Ribeiro}}
\email{caiocesarribeiro@alumni.usp.br}
\affiliation{International Center of Physics, Institute of Physics, University of Brasilia, 70297-400 Brasilia, Federal District, Brazil} 
\begin{abstract}
Nonlinear electrodynamics has been an important area of research for a long time. Investigations based on nonlinear Lagrangians, such as Euler-Heisenberg and Born-Infeld, are instrumental in exploring the limits of classical and quantum field theories, providing valuable insights into strong-field phenomena. In this context, this work considers how light propagates in strong-field environments, where such nonlinearities play significant roles, offering a way to investigate events in high-energy astrophysics, quantum optics, and fundamental physics beyond classical Maxwell's framework. Here, several aspects of light propagation in nonlinear electrodynamics are discussed. Phase and group velocities are derived and several interesting behaviors are unveiled, such as birefringence, non-reciprocal propagation, and asymmetries between phase and group velocities in special configurations. Specific solutions based on commonly studied nonlinear theories are also investigated, and phenomena like slow-light and one-way propagation are discussed. 
\end{abstract}

\maketitle

\setlength{\parskip}{0pt}
\section{Introduction}
\label{one}

When a light ray propagates in the presence of a strong electromagnetic field, its velocity depends on its wave polarization. In the regime of electromagnetic fields that are comparable to the Schwinger critical fields $B_{cr}= m_e^2 c^2/(e \hbar)\approx 4.41\times10^{9}$T or $E_{cr}=c B_{cr}$, Maxwell's electrodynamics should be amended by corrections of nonlinear nature, as described by the quantum electrodynamics (QED) \cite{1936ZPhy...98..714H,Euler:1935zz,weisskopf1936elektrodynamik,PhysRev.82.664}. Bialynicka-Birula and Bialynicki-Birula \cite{PhysRevD.2.2341}, among many others (for a recent review, see \cite{https://doi.org/10.1002/prop.202200092} and references therein), have studied the effects of nonlinearities in the propagation of electromagnetic waves. 

Some of the most prevalent nonlinear theories are Born-Infeld \cite{1933Natur.132.1004B,1934RSPSA.144..425B}, Euler-Heisenberg \cite{1936ZPhy...98..714H,Euler:1935zz,weisskopf1936elektrodynamik}, and the recently reported ModMax \cite{2020PhRvD.102l1703B,KOSYAKOV2020135840}. The Born-Infeld theory proposes the existence of a reference field  \cite{fouche2016limits} defined as the upper limit of a purely electric field, of $1.186\times 10^{20}{\rm V}{\rm m}^{-1}$ ($3.955\times 10^{11}$T, in units of $c=1$), which is obtained by equating the classical self-energy of the electron to its rest mass energy. In other words, it roughly corresponds to the electric field amplitude produced by an electron at a distance equal to its classical radius.  The Euler-Heisenberg theory is derived from the principles of QED and incorporates vacuum polarization phenomena at one-loop level. It is applicable for electromagnetic fields that change slowly compared to the inverse of the electron mass. 
As is well known, in the absence of free sources, Maxwell’s electrodynamics exhibits conformal and dual invariances. In contrast, the Born-Infeld theory lacks conformal invariance, and the Euler-Heisenberg theory is neither conformally nor dual invariant. On the other hand, the ModMax theory fulfills these two symmetries \cite{2020PhRvD.102l1703B}. Wave propagation in ModMax electrodynamics under uniform electric and magnetic fields has been investigated \cite{PhysRevD.107.075019}, including the analysis of birefringence phenomena.

Recent studies of magnetars and high-intensity lasers suggest they are among the most promising environments to test the predictions of nonlinear electrodynamics, as the magnitude of the involved magnetic fields approach the thresholds predicted by QED. Magnetars are characterized by a surface magnetic field reaching around $10^{10}-10^{11}$ T, and evidences of strong field effects around these objects exist \cite{10.1093/mnras/stw2798,Pereira2018}. Also, terrestrial laboratories have achieved peak laser intensities of $10^{19}$Wcm$^{-2}$ \cite{PhysRevLett.106.105002}, corresponding to electric fields of approximately $10^{11}$Vm$^{-1}$. Other projects like ELI, XCELS, and multi-petawatts LASERs are expected to reach $10^{19}-10^{26}$Wcm$^{-2}$ \cite{Raynaud_2018,Vranic_2018,ELI2024, XCELS2023}. Research in nonlinear electrodynamics has explored phenomena related to light-by-light scattering \cite{BAUR2002359, 10.1063/5.0150790}, a process that could shed some light on fundamental questions in particle physics. It has also been applied to describe certain black hole-like solutions \cite{Guzman-Herrera_2024,PhysRevD.97.084058}.

Nonlinear electrodynamics theories model vacuum polarization as an effective material medium, and a correspondence between the classical Maxwell's theory in a nonlinear dielectrics and a nonlinear theory of electromagnetism in vacuum exist \cite{PhysRevA.89.043822}. This connection opens a window for investigating analog models that connect these two important areas of research.

In the regime of geometrical optics, the propagation of photons in the presence of an intense external field is described as the propagation of weak perturbations. The corresponding wave equation is linear in the perturbed fields, but its coefficients can depend on the magnitude of the external field and also on the derivatives of the Lagrangian density with respect to the field invariants \cite{PhysRevD.2.2341}.  A light pulse can be built up by superposing plane wave solutions with different wave vectors, and it propagates with a velocity determined by its group velocity \cite{landau2013electrodynamics}. 

Here, wave propagation in nonlinear electrodynamics is investigated in the regime of linear optics. Employing a methodology typically used in optics in instantaneous media, general expressions for phase and group velocities are derived for arbitrary nonlinear Lagrangian models. In the special case where the nonlinear contributions are assumed as small perturbations, a first-order approximation shows that the magnitude of these velocities will always coincide, no matter the nonlinear theory considered, but their direction of propagation will be generally non-coincident. This result is useful if we want to compare nonlinear theories with experiments or observations that can measure wave-packed signals under high-intensity electromagnetic fields. Still, in the general framework, the behavior of the group velocity is examined in some distinct configurations. It is also unveiled that when wave propagation is set in the same plane containing the non-coincident external electric and magnetic fields, the group velocity will detach from this plane. In other words, if the wave vector is set in any direction in the plane spanned by the electric and magnetic fields, the group velocity will present components not only in that plane but also in the direction orthogonal to it. This seems to be a feature that has not yet been announced in the context of nonlinear electrodynamics. In addition, several results  are discussed in specific nonlinear models. 

The work is organized as follows. In the next section, the field equations for a general Lagrangian dependent on the two Lorentz and gauge invariants $F$ and $G$ are established. Section \ref{sect:disprel} deals with wave propagation in the context of a general nonlinear theory. The dispersion relation for the electromagnetic waves is derived, recovering early results. It is shown that all information about the considered nonlinear theory can be encompassed in the coefficients ($\xi_{\scriptscriptstyle \pm}$). Phase and group velocities are derived in Secs. \ref{sect:phasevel}  and \ref{sec:group}, respectively. In particular, it is shown that, no matter the nonlinear theory considered, both velocities coincide in the regime of small $\xi_\pm$. Additionally, it is shown that solutions that exhibit a group velocity outside the plane where the wave vector is defined are possible. In section \ref{sect:app} the results are applied to Born-Infeld, Euler-Heisenberg, and ModMax nonlinear theories. Implications and consequences are discussed in final remarks, Sec. \ref{final}.

The background spacetime is assumed to be the Minkowski one and is represented by $\eta_{\mu\nu}$, which is defined by $\rm{diag}(+1,-1,-1,-1)$. We set the units $c=1=\hbar$, but in some cases it is convenient to use SI units, where we express quantities in terms of volts (V) instead of kilograms (kg), as this choice is more directly relevant in the context of electrodynamics.

\section{Nonlinear spin-one theories}
\label{sect:spint}

The electromagnetic field strength is here represented by the anti-symmetric 
rank-two tensor $F_{\mu\nu}$, and its dual is defined as $F^{*}_{\alpha\beta} = \frac12\eta_{\alpha\beta}{}^{\sigma\tau}F_{\sigma\tau}$, where we defined the completely anti-symmetric tensor $\eta_{\alpha\beta\mu\nu}$ such that, in a Cartesian coordinate system, $\eta^{0123} = 1$. 

We start by assuming a general gauge invariant Lagrangian density for electrodynamics as an arbitrary function of the only two Lorentz invariant scalar fields, i.e. $L = L(F,G)$, where $F=F^{\mu\nu}F_{\mu\nu}$ and $G=F^{\mu\nu}F^*_{\mu\nu}$.

The least action principle leads to the field equation
\begin{equation}
\label{9}
\left(L_{\scriptscriptstyle F} F^{\mu\nu} + L_{\scriptscriptstyle G} F^{*\mu\nu}\right){}_{,\nu} = 0,
\end{equation}
where a comma denotes partial derivatives with respect to the Cartesian coordinates and
$L_{X}=\partial L/\partial X$, where $X$ stands for any monomial on the field invariants.  
Using relations
\begin{math}
F_{,\nu} = 2F^{\alpha\beta}F_{\alpha\beta,\nu} 
\end{math} and 
\begin{math}
G_{,\nu} = 2F^{\alpha\beta}F^*_{\alpha\beta,\nu} 
\end{math}
in Eq.~(\ref{9}), we obtain \cite{2000PhLB..482..134D}
\begin{equation}
\label{20}
2N^{\mu\nu\alpha\beta}F_{\alpha\beta ,\nu} 
+ L_{\scriptscriptstyle F} F^{\mu\nu}{}_{,\nu} = 0,
\end{equation} 
where we introduced the rank-four tensor
$N^{\mu\nu\alpha\beta}$ through 
\begin{align}
N^{\mu\nu\alpha\beta} \doteq {} &
L_{\scriptscriptstyle FF}F^{\mu\nu}F^{\alpha\beta} 
+ L_{\scriptscriptstyle GG} F^{*\mu\nu}F^{*\alpha\beta}
\nonumber \\
&+ L_{\scriptscriptstyle FG}\left( 
F^{\mu\nu}F^{*\alpha\beta} +
F^{*\mu\nu}F^{\alpha\beta}\right).
\label{21}
\end{align}
Notice that $N^{\mu\nu\alpha\beta}$ has some of the properties of the Riemann tensor, namely, $N^{\mu\nu\alpha\beta}=N^{\alpha\beta\mu\nu}$ and $N^{\mu\nu\alpha\beta}=-N^{\nu\mu\alpha\beta}$.
Additionally, the field strength $F_{\mu\nu}$ must satisfy the 
Bianchi identity: 
\begin{align}
F_{\mu\nu,\gamma}+F_{\nu\gamma,\mu}+F_{\gamma\mu,\nu}=0.
\label{bianchi}
\end{align}

In terms of electric and magnetic fields $\vec E$ and $\vec B$, and the auxiliary fields
\begin{align}
\vec D = -4(L_F\vec E + L_G \vec B),
\nonumber \\
\vec H= -4(L_F\vec B - L_G \vec E),
\nonumber
\end{align}
the field equations given in Eqs.~(\ref{20}) and (\ref{bianchi}) reduce to $\nabla \cdot \vec D = 0$, $\nabla \cdot \vec B = 0$, $\nabla\times\vec E = -\partial_t \vec B$, and $\nabla\times\vec H = \partial_t \vec D$. Once the Lagrangian is set in terms of the field invariants $F$ and $G$, the above expressions for the auxiliary fields provide the effective constitutive relations for the nonlinear theory.

\section{dispersion relations}
\label{sect:disprel}
Monochromatic plane-wave solutions satisfying the above field equations for the nonlinear electrodynamics can be obtained in the regime of linear optics as follows. The total electromagnetic field $F_{\mu\nu}$ is assumed to be split into two contributions, a strong and nearly constant background field $F^b_{\mu\nu}$ plus a relatively small but rapidly varying wave contribution $F^\omega_{\mu\nu} = f_{\mu\nu}\,\exp(i k_\alpha x^\alpha)$, where $k_\alpha = (\omega,-\vec q)$ is the wave four-vector whose components are the frequency $\omega$ of the wave and its wave vector $\vec q$, and the amplitude tensor field $f_{\mu\nu}$ is related to the polarization vector ${\rm e}_\mu$ of the wave by means of \cite{2000PhLB..482..134D} $f_{\mu\nu} \propto {\rm e}_\mu k_\nu - {\rm e}_\nu k_\mu$. By employing this prescription in the field equations, it follows that the wave field $F^\omega_{\mu\nu}$ obeys a linear wave equation, whose coefficients depend on the background field $F^b_{\mu\nu}$, and also on the specific nonlinear theory under consideration through the derivatives of the corresponding Lagrangian density. The eigenvalue problem $Z_{ij}e_j = 0$ follows from this approach, where it was defined
\begin{align}
Z^{\mu\nu} = I^{\mu\nu} +\frac{4}{L_{\scriptscriptstyle F} k^2} N^{\mu\alpha\nu\beta}k_\alpha k_\beta,
\label{Zmn}
\end{align}
where we have introduced the orthogonal projector $I^{\mu\nu}= \eta^{\mu\nu} -k^\mu k^\nu/k^2$. 

It is helpful to observe that $I^{\mu\nu}k_\nu = 0$ and due to the symmetries in the $N^{\mu\alpha\nu\beta}$ tensor, it follows that 
\begin{align}
Z^{\mu\nu}k_\nu = 0.
\label{DetZ}
\end{align}
This piece of information is valuable when calculating the determinant of this tensor by means of its traces: $6\det(Z_{\mu\nu}) = (Z_1)^3 -3 Z_1 Z_2 + 2 Z_3$, where $Z_{1}=Z^\mu{}_\mu$, $Z_{2} = Z^{\mu}{}_{\nu}Z^{\nu}{}_{\mu}$, and $Z_{3} = Z^{\alpha}{}_{\nu}Z^{\nu}{}_{\mu}Z^{\mu}{}_{\alpha}$.

Notice that, as $\| F^b_{\mu\nu} \| >> \|F^\omega_{\mu\nu}\|$, all electromagnetic fields appearing in Eq.~(\ref{Zmn}) can be approximated to $F^b_{\mu\nu}$ only. To maintain a simple notation, the superscript $b$ in those fields will be omitted hereafter. 

Solutions for this eigenvalue problem in Eq.~(\ref{DetZ}) can be found by solving $\det(Z_{\mu\nu}) = 0$ for $k^2$, and results in the dispersion relation \cite{2000PhLB..482..134D}:
\begin{eqnarray}
k^2=\xi_{\scriptscriptstyle \pm}
F^{\mu\sigma}F_{\sigma}{}^{\nu}k_{\mu}k_{\nu},
\label{57}
\end{eqnarray}
where we defined
\begin{align}
\xi_{\scriptscriptstyle \pm} \doteq 4\frac{L_{\scriptscriptstyle FF}+ \Omega_{{\scriptscriptstyle \pm}}  
L_{\scriptscriptstyle FG}}{L_{\scriptscriptstyle F} + G\left(L_{\scriptscriptstyle FG}+
\Omega_{{\scriptscriptstyle \pm}}L_{\scriptscriptstyle GG}\right)},
\end{align}
with
\begin{equation}
\Omega_{{\scriptscriptstyle \pm}} = \frac{-\Omega_2 \pm \sqrt{\Omega_2^2 - 4\Omega_1\Omega_3}}{2\Omega_1},
\label{55}
\end{equation}
and also 
$\Omega_1 \doteq {} -L_{\scriptscriptstyle FG}(L_{\scriptscriptstyle F} -2FL_{\scriptscriptstyle GG}) - G\left(L_{\scriptscriptstyle FG}^2 - L_{\scriptscriptstyle GG}^2\right)$, 
$\Omega_2 \doteq {} 2F\left(L_{\scriptscriptstyle FF}L_{\scriptscriptstyle GG} + L_{\scriptscriptstyle FG}^2\right) + (L_{\scriptscriptstyle F}+2GL_{\scriptscriptstyle FG})(L_{\scriptscriptstyle GG}-L_{\scriptscriptstyle FF})$, 
and 
$\Omega_3 \doteq {} L_{\scriptscriptstyle FG}(L_{\scriptscriptstyle F} +2FL_{\scriptscriptstyle FF}) - G(L_{\scriptscriptstyle FF}^2 -L_{\scriptscriptstyle FG}^2)$.

It is interesting to mention that $k^2=0$ is a possible solution in nonlinear theories, provided $L_{\scriptscriptstyle FF} + \Omega_{{\scriptscriptstyle \pm}}L_{\scriptscriptstyle FG}=0$.  
As already noticed  \cite{2000PhRvD..61d5001N} a family of Lagrangians satisfying this condition is $L = -F/4 + f(G)$, with $f(G)$ an arbitrary function 
of $G$.

\section{phase velocity}
\label{sect:phasevel}
Since $k^2 = \omega^2 - q^2$, where $q^2 = \vec q \cdot \vec q= -k^i k_i$, the phase velocity
$\omega/q\equiv v$ of a propagating wave is
found to be
\begin{eqnarray}
v^2_{\scriptscriptstyle \pm}=1+\xi_{\scriptscriptstyle \pm}
F^{\alpha\sigma}F_{\sigma}{}^{\beta}n_{\alpha}n_{\beta},
\label{phasepn}
\end{eqnarray}
where we defined $n_\mu \doteq k_\mu/q$. 

Notice that, once $n_0 = v$, the above equation consists in a quadratic equation for the phase velocity which, in its standard form, reads
\begin{align}
&(1-\xi_{\scriptscriptstyle \pm} F^{0i}F_{i}{}^0)v^2 - 2 \xi_{\scriptscriptstyle \pm} F^{0i}F_{i}{}^j n_j v 
\nonumber \\ 
&- (1+ \xi_{\scriptscriptstyle \pm} F^{i\alpha}F_{\alpha}{}^j n_i n_j)=0,
\label{phaseequation}
\end{align}
whose solutions are given by
\begin{align}
v_{\scriptscriptstyle \pm} =& \Big[1-\xi_{\scriptscriptstyle \pm} F^{0i}F_{i}{}^0\Big]^{-1}\bigg\{\xi_{\scriptscriptstyle \pm} F^{0i}F_{i}{}^j n_j \pm \Big[(\xi_{\scriptscriptstyle \pm} F^{0i}F_{i}{}^j n_j)^2
\nonumber \\
&+(1-\xi_{\scriptscriptstyle \pm} F^{0i}F_{i}{}^0)(1+ \xi_{\scriptscriptstyle \pm} F^{i\lambda}F_{\lambda}{}^j n_i n_j)\Big]^{1/2}\bigg\}.
\label{phasev}
\end{align}
The $\pm$ sub-indices in $\xi_{\scriptscriptstyle \pm}$ are independent of the $\pm$ signs in the remaining equation, resulting in a maximum of four independent solutions for the phase velocity. To be more specific, we will generally have two solutions $v_+ = v_+(\xi_\pm)$ and two solutions  $v_- = v_-(\xi_\pm)$.
A direct inspection shows that these possible solutions lead to phase velocities that generally depend on the direction of propagation. Birefringence phenomena can appear in any direction, depending on $\xi_{\scriptscriptstyle \pm}$ taking distinct values for a given theory, and space-reversal symmetry will be broken if $F^{0i}F_{i}{}^j n_j \neq 0$. 

It is worth presenting the possible solutions for Eq.~(\ref{phaseequation}) in the regime of small magnitude of $\xi$ (as compared to the inverse of the squared electromagnetic fields). In such a regime, we obtain
\begin{equation}
v = \left\{ 
\begin{array}{c}
1 + \frac{\xi_{\scriptscriptstyle +}}{2}\left(F^{0i}F_{i}{}^0 + 2F^{0j}F_{j}{}^i n_i+ F^{i\alpha}F_{\alpha}{}^j n_i n_j \right), \\[1ex]
1 + \frac{\xi_{\scriptscriptstyle -}}{2}\left(F^{0i}F_{i}{}^0 + 2F^{0j}F_{j}{}^i n_i+ F^{i\alpha}F_{\alpha}{}^j n_i n_j \right), \\[1ex]
-1 - \frac{\xi_{\scriptscriptstyle +}}{2}\left(F^{0i}F_{i}{}^0 - 2F^{0j}F_{j}{}^i n_i+ F^{i\alpha}F_{\alpha}{}^j n_i n_j \right), \\[1ex]
-1 - \frac{\xi_{\scriptscriptstyle -}}{2}\left(F^{0i}F_{i}{}^0 - 2F^{0j}F_{j}{}^i n_i+ F^{i\alpha}F_{\alpha}{}^j n_i n_j \right),
\end{array} \right. 
\label{approxv}
\end{equation}
which confirms, for this special regime, the anticipated behavior of birefringence and non-reciprocal propagation. 

\section{group velocity}
\label{sec:group}
The group velocity can be obtained by means of the definition $\vec u = (\partial \omega /\partial\vec q)$, which, in component notation, would read $u^i = (\partial \omega /\partial k_i)$. Thus, as $\omega = q v$ we obtain
\begin{align}
    u^i = \frac{\partial\omega}{\partial k_i} = v\frac{\partial q}{\partial k_i} + q\frac{\partial v}{\partial k_i}.
    \label{groupui}
\end{align}
Observing that $q = \sqrt{-\eta^{ij}k_ik_j}$, it follows that $(\partial q /\partial k_i) = -k^i/q$, and $u^i = -v (k^i/q) + q (\partial v /\partial k_i)$. 

Finally, taking the derivative of Eq.~(\ref{phasepn}),
\begin{align}
    \frac{\partial v_{\scriptscriptstyle \pm}^2}{\partial k_i} = \frac{2\xi_{\scriptscriptstyle \pm}}{q^2}F^{\alpha\sigma}k_\alpha\left( F_{\sigma}{}^{0} u^i  +F_{\sigma}{}^{i} + F_{\sigma}{}^{\beta}n_\beta n^i \right),
\end{align}
and using this result in Eq.~(\ref{groupui}), after some algebra, we obtain that  
\begin{align}
    u^i_{\scriptscriptstyle \pm} = \frac{-v_{\scriptscriptstyle \pm}^2 n^i + \xi_{\scriptscriptstyle \pm}\left( F^{\alpha\sigma}F_{\sigma}{}^{i}n_\alpha + F^{\alpha\sigma}F_{\sigma}{}^{\beta}n_\alpha n_\beta n^i \right)}{v_{\scriptscriptstyle \pm} -\xi_{\scriptscriptstyle \pm}F^{\mu\nu}F_{\nu}{}^{0}n_\mu},
    \label{group1}
\end{align}
or yet, conveniently reintroducing $v_{\scriptscriptstyle \pm}^2$ in this result, the group velocity can be more compactly expressed as
\begin{align}
    u^i_{\scriptscriptstyle \pm} = \frac{- n^i + \xi_{\scriptscriptstyle \pm} F^{\alpha\sigma}F_{\sigma}{}^{i}n_\alpha}{ {v_{\scriptscriptstyle \pm}} -{\xi_{\scriptscriptstyle \pm}}F^{\mu\nu}F_{\nu}{}^{0}n_\mu}.
\label{group2}
\end{align}

An interesting result emerges from this expression when we assume that $\xi$ is a small perturbation. In a first-order approximation, it can be shown that the magnitudes of the phase and group velocities coincide, i.e., $$u^2 = -u^iu_i= v^2 + {\cal O}(\xi^2).$$ Thus, no matter the nonlinear theory considered, in this regime, the wave packet will travel with the same velocity of plane waves. Notice, however, that their directions are generally non-coincident. 

It is worth mentioning that a 4-vector notation for the group velocity is possible using the orthogonal projector $h_{\mu\nu} = \eta_{\mu\nu} -V_\mu V_\nu$, where $V_\mu$ denotes the 4-velocity field of an observer. Usually, it is convenient to assume that the observer is co-moving with the system, which would lead us to set $V_\mu = \delta_\mu^0$. In such a description, the group velocity could be defined as $u^\mu = h^{\mu\nu}(\partial\omega/\partial k^\nu)$.

Let us now derive the group-velocity components in order to obtain a description for the direction of propagation of wave packets in nonlinear electrodynamics. We choose a field configuration such that the only non-null components of the electromagnetic tensor field are given by $F_{\mu\nu} = E(\delta_{0\mu}\delta_{1\nu}-\delta_{1\mu}\delta_{0\nu})+B(\delta_{1\mu}\delta_{3\nu}-\delta_{3\mu}\delta_{1\nu})$, i.e. the electric field is aligned in the $x$ direction and the magnetic field is aligned in the $y$ direction. The components of the 4-vector $n_\mu$, that gives the wave-vector direction, in spherical coordinates $(r,\theta,\varphi)$ are expressed as $n_{\mu}=(v,-\sin\theta \cos\varphi,- \sin\theta \sin\varphi,-\cos\theta)$.  
Thus, the Cartesian components are:
\begin{align}
    u_x^{\scriptscriptstyle \pm} = & \frac{1-\xi_{\scriptscriptstyle \pm} (E^2-B^2)}{(1-\xi_{\scriptscriptstyle \pm} E^2)v_{\scriptscriptstyle \pm} + \xi_{\scriptscriptstyle \pm} EB \cos\theta} \sin\theta\cos\varphi,
\label{groupx}
\\
u_y^{\scriptscriptstyle \pm} = & \frac{\sin\theta\sin\varphi}{(1-\xi_{\scriptscriptstyle \pm} E^2)v_{\scriptscriptstyle \pm} +  \xi_{\scriptscriptstyle \pm} EB \cos\theta},
\label{groupy}
\\
u_z^{\scriptscriptstyle \pm} = & \frac{(1+\xi_{\scriptscriptstyle \pm}B^2)\cos\theta -  \xi_{\scriptscriptstyle \pm} EBv_{\scriptscriptstyle \pm}}{(1-\xi_{\scriptscriptstyle \pm} E^2)v_{\scriptscriptstyle \pm} + \xi_{\scriptscriptstyle \pm} EB \cos\theta}.
\label{groupz}
\end{align}

Let us examine closer the solutions for propagation in the three different orthogonal planes.

\subsection{Propagation with the wave vector in the $xz$ plane }
Choosing $\varphi=0$, which corresponds to a wave vector $\vec n = (\sin\theta,0,\cos\theta)$, we obtain that the group velocity will propagate in the same plane of $\vec q$, but with a different direction. Using the above results, it can be shown that the angle $\psi$ the group velocity does with the $z$ axis is given by 
\begin{align}
   \psi = \arctan \frac{\big[1-\xi_{\scriptscriptstyle \pm} (E^2-B^2)\big]\sin\theta}{(1+\xi_{\scriptscriptstyle \pm}B^2)\cos\theta - \xi_{\scriptscriptstyle \pm} EBv_{\scriptscriptstyle \pm}}.
\label{psifi0xz}
\end{align}
In this configuration, the group and phase velocities propagate in different directions, forming an angle 
$|\psi-\theta|$ between them. Notice that, when $\xi_{\scriptscriptstyle \pm} \to 0$, these directions coincide, as expected.

\subsection{Propagation with the wave vector in the $yz$ plane }
Now, by choosing $\varphi=\pi/2$, which corresponds to a wave vector $\vec n = (0,\sin\theta,\cos\theta)$ we obtain a similar result, but now the angle $\psi$ the group velocity does with the $z$ axis is given by 
\begin{align}
   \psi = \arctan \frac{\sin\theta}{(1+\xi_{\scriptscriptstyle \pm}B^2)\cos\theta - \xi_{\scriptscriptstyle \pm} EBv_{\scriptscriptstyle \pm}}.
\label{psifi0yz}
\end{align}
\subsection{Propagation with the wave vector in the $xy$ plane }

Finally, by choosing $\theta=\pi/2$, which corresponds to a wave vector $\vec n = (\cos\varphi,\sin\varphi,0)$, we obtain that the group velocity will present three non-null components, i.e., there will be a component outside the plane $xy$, as can be inferred directly from Eqs.~(\ref{groupx}) - (\ref{groupz}). This is quite interesting behavior if we recall that these are vacuum solutions. This kind of behavior is expected to occur in the context of wave propagation in dielectric media. This effect seems to open a window to investigate analog models between vacuum nonlinear electrodynamics and electrodynamics in material media. Some steps in this direction were already discussed in the case of Born-Infeld-like solutions \cite{PhysRevA.89.043822}. 

The cosine of the angle between the group and phase velocities is given by
\begin{align} 
 \frac{1 - \xi_{\scriptscriptstyle \pm} (E^2 - B^2)\cos^2\varphi}{\sqrt{1 + \xi_{\scriptscriptstyle \pm}^2 E^2 B^2 v_{\scriptscriptstyle \pm}^2 +[\xi_{\scriptscriptstyle \pm}^2(E^2-B^2) - 2 \xi_{\scriptscriptstyle \pm}](E^2-B^2)\cos^2\varphi}}.
\nonumber
\end{align}
Notice that, in the regime of small $\xi_{\scriptscriptstyle \pm}$ the above result reduces to $1 + {\mathcal O}(\xi_{\scriptscriptstyle \pm})^2$, as anticipated.

\section{Applications}
\label{sect:app}
In what follows, we examine three well-known nonlinear models for electrodynamics. We set the field configuration $F_{01}=-F_{10}=E$ and $F_{13}=-F_{31} = B$, and restrict the analysis to wave propagation in $xz$-plane, i.e., $\varphi =0$. Thus, the unit vector $n_i$ in spherical coordinates is hereafter $n_i = (-\sin\theta,0,-\cos\theta)$. Furthermore, to simplify the description, we extend the range of variable $\theta$ from $0$ to $2\pi$, which allows us to present the results using fewer equations.

\subsection{Euler-Heisenberg Lagrangian}
\label{EHLsection}
In the limit of weak-field approximation, Euler-Heisenberg \cite{1936ZPhy...98..714H,PhysRev.82.664} Lagrangian density describing quantum-electrodynamics reads
\begin{equation}
L_{QED} = -\frac{1}{4}F + \frac{\mu}{4}\left(F^2 + \frac{7}{4}G^2\right),
\label{euler}
\end{equation}
where the $\hbar$-order parameter is
$\mu \doteq {2\alpha^2}/({45 m_{\scriptscriptstyle e}^4})=1.32 \times 10^{-24}{\rm T}^{-2}$.
Since $\mu F$ is a very small quantity, the solutions for the phase velocity can be obtained directly from Eq.~(\ref{approxv}), where we observe $\xi_{\scriptscriptstyle +}=-8\mu$ and $\xi_{\scriptscriptstyle -}=-14\mu$. Thus, it can be concluded that birefringence and non-reciprocal propagation are also present in Euler-Heisenberg electromagnetism, the latter occurring when a magnetic field is present. Because of the weak field approximation in Eq.~(\ref{euler}), the discussion following Eq.~(\ref{group2}) allows us to infer that the same results are immediately extended to the group velocity, which is thus identified with the phase velocity. 

\begin{figure}[b]
    \includegraphics[scale=0.55]{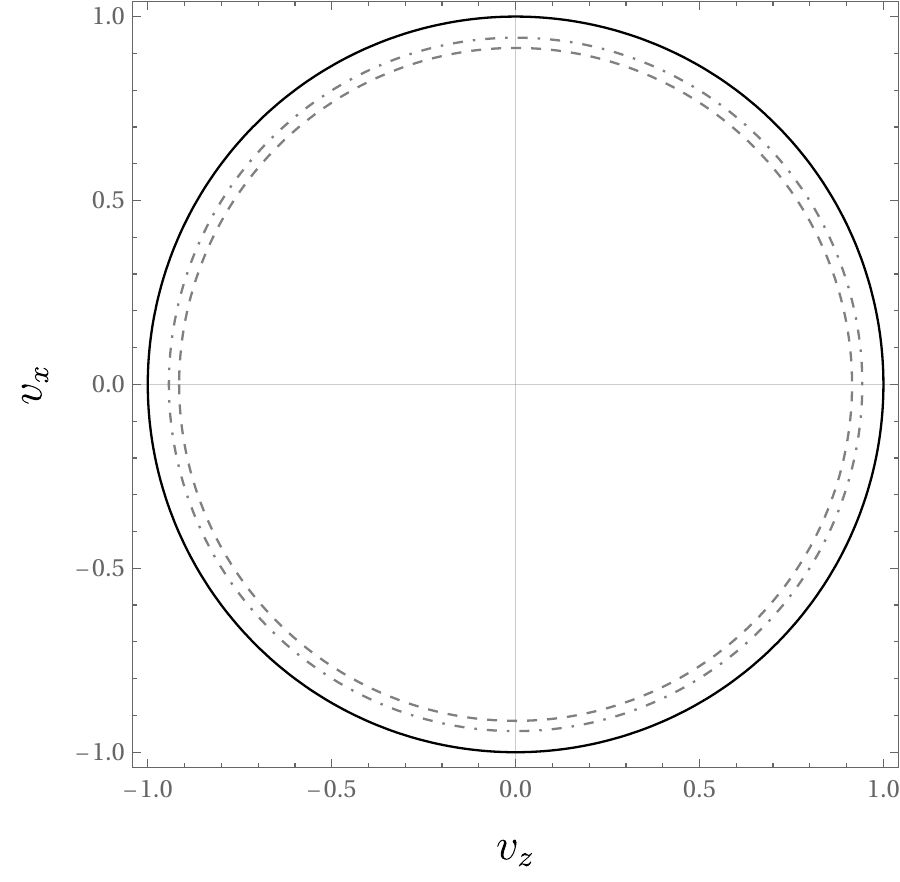}
    \caption{Normal surfaces for wave propagation in $xz$ plane in the Euler-Heisenberg model. Here, the field strengths were chosen to meet the expected values in a typical magnetar, i.e. $E=10^{13}$Vm$^{-1}$ and $B=10^{11}$T (in SI units). Despite the occurrence of birefringence and non-reciprocal propagation, only the former is visible in the figure. The external solid curve was added for comparison and represents a normal surface of light predicted by the linear theory.}
    \label{figEH1}
\end{figure}
The four solutions described by Eqs.~(\ref{approxv}) are condensed into the two solutions below,
\begin{align}
v = \left\{ 
\begin{array}{c}
1 -4\mu\left(E \cos\theta - B\right)^2,
\\[1ex]
1 -7\mu\left(E \cos\theta - B\right)^2 ,
\end{array} \right. 
\label{approxEBset}
\end{align}
where now $\theta$ varies from $0$ to $2\pi$.
Notice that the term proportional to the product of $EB$ is the only one responsible for the non-reciprocal propagation, which thus depends on the presence of both fields at the same time. 
In Fig.~\ref{figEH1} the normal surfaces \cite{1999poet.book.....B} for wave propagation in the $xz$ plane are depicted using electromagnetic field intensities typically reported in certain astrophysical systems, for instance, in magnetars, where the magnetic field at its surface achieves $10^{11}{\rm T}$ \cite{2005Natur.434.1107P,2022ApJ...926..111W,2013Natur.500..312T,2021Univ....7..351I}. The distance from the origin to a given point of a curve gives the magnitude of the phase velocity in such a direction.
The high values of these fields can produce a noticeable effect on light propagation, which is beyond the predictions of the linear theory. 
\begin{figure}[b]
    \includegraphics[scale=0.67]{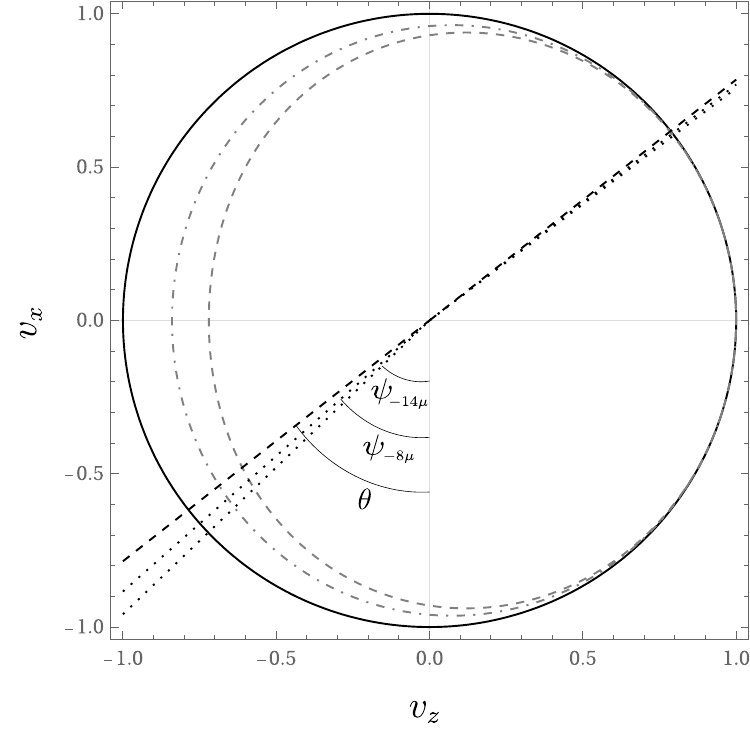}
    \caption{Normal surfaces for wave propagation in $xz$ plane in the context of Euler-Heisenberg model. The fields were set to $\mu E^2 \sim \mu B^2 \approx 0.01$.  At this scale, birefringence and non-reciprocal propagation phenomena are visible in the figure. Like in Fig.~\ref{figEH1}, the normal surface (the external solid circle) for light propagation in the linear theory ($v=1$) was included for the sake of comparison.}
    \label{figEH2}
\end{figure}
Notice that the birefringence effect occurs in any direction of propagation in this plane ($\varphi =0$). Despite the curves appearing as perfect circles in this scale, the propagation is non-reciprocal. To visualize such a distortion, we use Eq. (\ref{approxEBset}) and choose the electromagnetic fields $\mu E^2 \sim \mu B^2 \approx 10^{-2}$ in Fig.~\ref{figEH2}. Although the depicted value of the electric field in this figure may be exaggerated for illustrative purposes, extreme events capable of generating such high electric fields, as predicted in gamma-ray bursts, remain plausible. In this figure, we see that propagation in opposite directions, indicated by the straight dashed and dotted lines, is not symmetric. Additionally, this figure also unveils the fact that for a given direction of the wave vector, represented by the dashed straight line, there will be different directions for the group velocities, depicted by the dotted straight lines, which are determined by the angles $\psi_{-14\mu} = \psi(\xi=-14\mu)$ and $\psi_{-8\mu}= \psi(\xi=-8\mu)$.

\subsection{Born-Infeld Theory}
\label{sec-BI}
The Lagrangian density describing Born-Infeld electrodynamics is given by \cite{1934RSPSA.144..425B} 
\begin{align}
    L_{BI} =\beta^2\left( 1 - \sqrt{1 + \frac{F}{2\beta^2} - \frac{G^2}{16\beta^4}}\right), \label{BI}
\end{align}
where $\beta$ is a dimensional constant that defines a reference magnitude for the electromagnetic field,  which is estimated to be of the order of $4\times 10^{11}{\;}{\rm T} $ \cite{1934RSPSA.144..425B}. In this theory, the electric field  is restricted by the relation 
\begin{align} 
E \leq \beta\sqrt{\frac{\beta^2 + B^2}{\beta^2 + B^2 \cos^2\zeta }},
\label{conditionBI}
\end{align}
where $\zeta$ is the angle between the electric and the magnetic fields (for works exploring the limits of $\beta$ see Ref.~\cite{BialynickiBirula1983}). 
A simple solution showing this possibility can be easily obtained in the regime of uniform fields. For example, assuming that the auxiliary fields $\vec D$ and $\vec H$ are constants and perpendicular to each other, it is straightforward to show that the electric and magnetic fields are given by
\begin{align}
    \vec E = \frac{\vec D}{\sqrt{1+\frac{D^2-H^2}{\beta^2}}},
    \nonumber\\
    \vec B = \frac{\vec H}{\sqrt{1+\frac{D^2-H^2}{\beta^2}}}.
    \nonumber
\end{align}
In such case, $\vec E$ and $\vec B$ are also constants and orthogonal fields and can assumed values larger than $\beta$ for certain values of $\vec D$ and $\vec H$. To show one possible configuration, setting the auxiliary field strengths $D \approx 2.3349\beta$ and $H \approx 1.7961\beta$, it follows that $E \approx 1.3000\beta$ and $B \approx 1.0000\beta$, which are consistent solutions for this nonlinear electrodynamics.

It is important to note that the two possible solutions of the nonlinear factor $\xi_{\pm}$ coincide, i.e., $\xi_+=\xi_- =\xi= -2/(F+2 \beta^2)$. This fact is usually taken as an explanation for the non-existence of birefringence in Born-Infeld theory \cite{PhysRevD.2.2341,10.1063/5.0150790}. However, we anticipate that under certain conditions, birefringence and one-way propagation may appear. 

Considering the same configuration for the fields and the wave vector as before, we obtain the following solutions for the phase velocities:
\begin{align}
\nonumber
v_{\scriptscriptstyle \pm} = \frac{1}{B^2+\beta ^2} \Big[ \pm \Big\{&(B^2+\beta ^2) \left[(E^2-B^2) \sin ^2\theta +\mathcal{J}_0 \right]
\\
&-B^2 \cos ^2\theta \mathcal{J}_0 \Big\}^{\frac{1}{2}} +  B E \cos\theta \Big],
\nonumber
\end{align}
where we have defined $\mathcal{J} \doteq B^2+\beta ^2-E^2 \cos^2\theta$, with $\mathcal{J}_0 = \mathcal{J}(\theta=0)$.

\begin{figure}[t]
    \includegraphics[width=\linewidth]{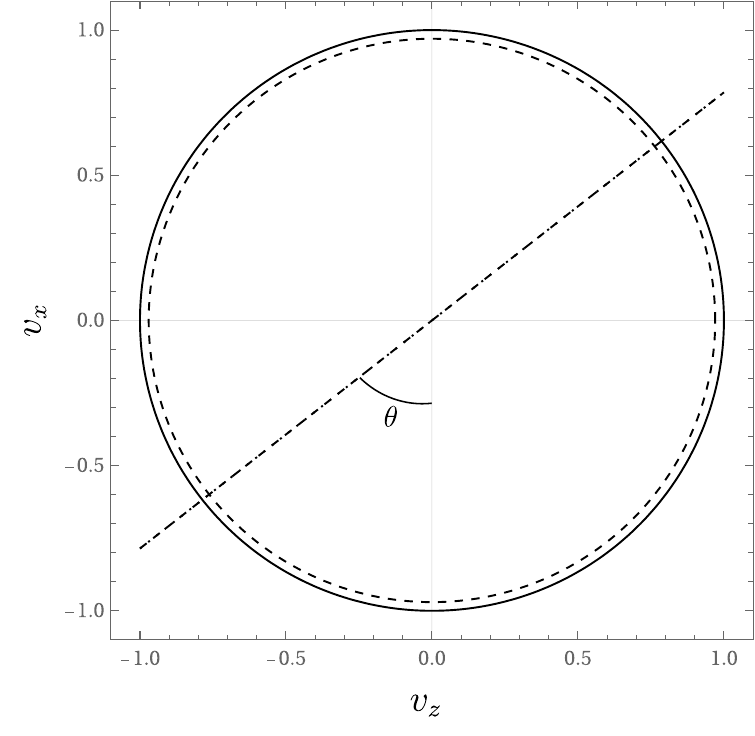}
    \caption{Normal surfaces for wave propagation in the plane $xz$ in Born-Infeld theory. The solid curve represents a normal surface of light in Maxwell's theory, while the dashed one represents the solution predicted by the nonlinear theory. The electromagnetic fields are set to $E=10^{13}$Vm$^{-1}$ and $B=10^{11}$T (in SI units), which are consistent with field strengths predicted to occur in certain magnetars. Non-reciprocity of propagation is present but is not visible in this plot. }
    \label{figBIMag}
\end{figure}
When the propagation is set in the plane $xz$ ($\varphi=0$), it follows that 
\begin{align}
    v_{\scriptscriptstyle \pm} = \frac{1}{B^2+\beta ^2}\Big( B E \cos\theta  \pm \beta\sqrt{\mathcal{J}}   \Big).
    \label{phasevelbixz}
\end{align}

Note that while Born-Infeld theory allows for only one value of the birefringence coefficient $\xi$, there will be non-reciprocal propagation in the presence of both electric and magnetic fields.

We now calculate the components of the group velocities, recalling Eqs. (\ref{groupx})-(\ref{groupz}), for propagation in the $xz$ plane,
\begin{align}
    u_z &=\frac{1}{B^2+\beta ^2} \left(B E\pm \frac{\beta \mathcal{J}_0 \cos\theta}{\sqrt{\mathcal{J}}} \right),
    \\
    u_x &= \pm \frac{\beta  \sin\theta}{\sqrt{\mathcal{J}}},
\end{align}
with the direction of propagation determined by 
\begin{align}
 \psi_{\pm} = \arctan \left(\pm\frac{ \beta  \left(B^2+\beta ^2\right) \sin \theta}{B E \sqrt{\mathcal{J}} \pm \beta \mathcal{J}_0  \cos\theta }\right).
\end{align}
These results are in agreement with the one reported in a previous publication \cite{PhysRevA.89.043822}. 
In Fig.~\ref{figBIMag} the normal surface for the phase velocity is depicted by the dashed curve, again considering electromagnetic fields whose intensities are consistent with dipole field strengths at the surface of certain magnetars \cite{2022ApJ...926..111W,2013Natur.500..312T,2021Univ....7..351I}. 
\begin{figure}[b]
    \includegraphics[width=\linewidth]{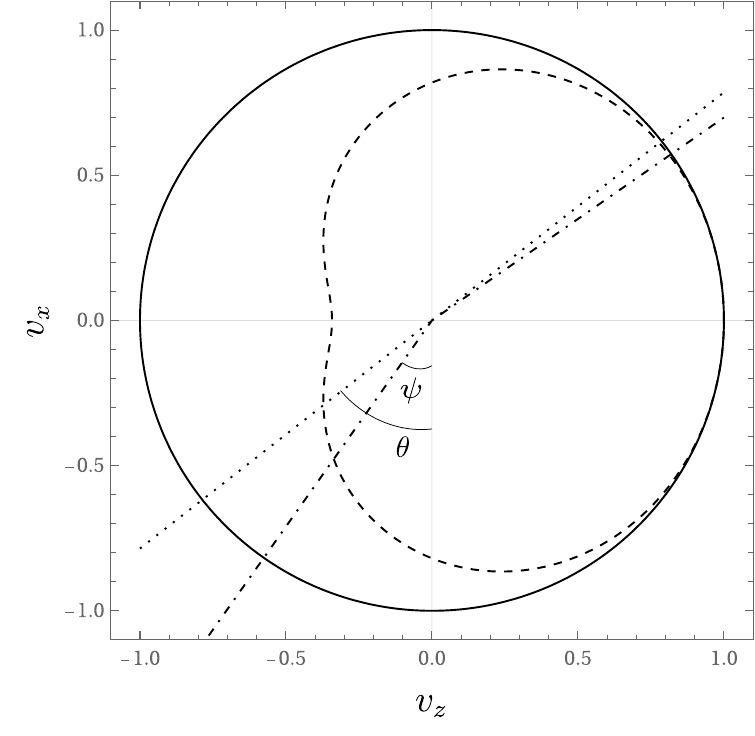}
    \caption{Normal surfaces for wave propagation in plane $xz$ in Born-Infeld theory. The dotted line represents the direction of propagation of phase velocity while the dotted-dashed line represents the direction of the group velocity. The electric and magnetic fields were set to $E=8.3\times 10^{19}$Vm$^{-1}$ and $B=2.8\times 10^{11}$T (in SI units), respectively. The solid curve represents a normal surface of light in Maxwell's theory, the non-reciprocity is illustrated by the dashed curve.}
    \label{figBIexa}
\end{figure}
Here, the non-reciprocity effect is not visually evident because of the high value of the critical field presented by the theory. However, the difference in the magnitude of the phase velocity as compared to Maxwell's prediction is clear. To make it possible to visualize the effects of non-reciprocal propagation, in Fig.~\ref{figBIexa} we have chosen the electromagnetic fields such that $E = B = 0.7 \beta$. In this figure, the difference between the directions of group and phase velocities is illustrated for a given wave vector. For example, for the wave vector represented by the dotted straight line, determined by the angle $\theta$, the corresponding group velocity will have the direction given by the dot-dashed line. Notice that the magnitude of this effect is not symmetric in opposite directions.

\begin{figure}[t]
    \includegraphics[width=\linewidth]{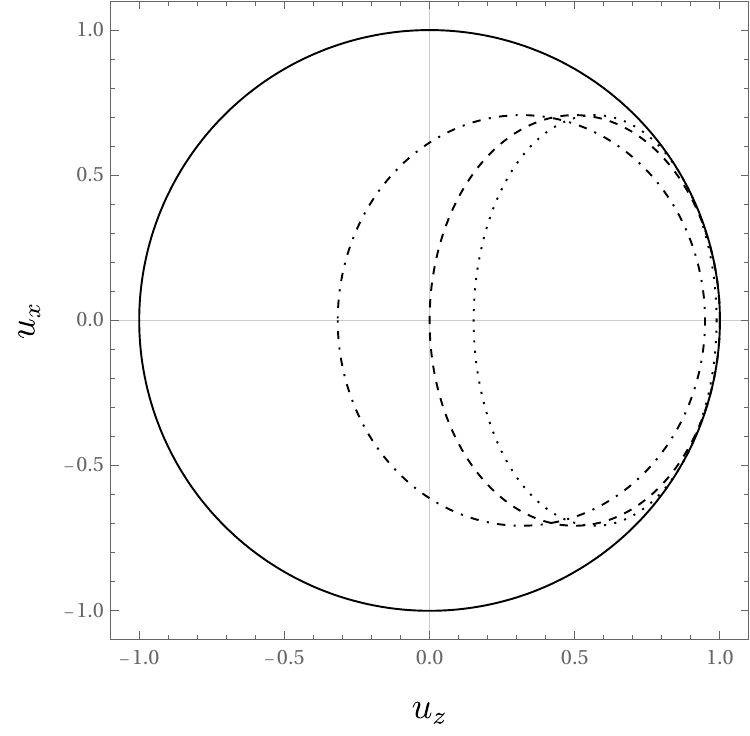}
    \caption{Normal surfaces for group velocity in plane $xz$ in Born-Infeld theory. The background magnetic field for all the curves is taken as $B = \beta$, and the values for the electric field were set to $E = 0.63 \beta$, $E = \beta$, and $E=1.14\beta$, corresponding to the dot-dashed, dashed, and dotted curves, respectively. Note that for the dashed curve, $E=B=\beta$, the magnitude of the group velocity is closer to zero for propagation sufficiently close to the x-direction. As $E$ becomes larger than $\beta$, the two solutions evolve in such a way that birefringence could be present in this regime.
    }
    \label{figBIgroupvel}
\end{figure}

A remarkable result in optics is the existence of slowlight and one-way propagation. The first one refers to a light speed closer to zero, and the second one means that both solutions for group velocity occur in the same direction. 
When we calculate the group velocity as $u^2=-u^i u_{i}$, we obtain two expressions that correspond to wave propagation in opposite directions, as expected in Born-Infeld electrodynamics (recap that, in such a theory, there exists only one solution for the birefringence coefficient $\xi$). In the same way as occurs in the phase velocity, non-reciprocal propagation is also present in the group velocity, as illustrated in Fig.~\ref{figBIgroupvel}. 
The slow light effect is only appreciated for values of the electric field closer to $\beta$. In Fig.~\ref{figBIgroupvel}, the group velocity is plotted for different values of the electric field, where the distance from the origin to a given point of a curve gives the magnitude of the group velocity in such a direction. When $E=0.63\beta$, the dot-dashed curve, the group velocity presents distinct solutions in opposite directions, which is a feature of non-reciprocal propagation. In this case, there is no birefringence, as expected. However, as we increase the value of $E$, the whole curve evolves to the right, eventually touching the origin at one point when $E=\beta$. In such a case, the system starts to exhibit one-way propagation: only propagation in the positive direction (with positive $u_z$ component) is allowed. Furthermore, if the electric field is allowed to be larger than $\beta$, there will still be one-way propagation, but now with two distinct solutions, as depicted by the closed dotted curve in this figure. In conclusion, when fields higher than $\beta$ are allowed, but still satisfying the condition stated by Eq.~(\ref{conditionBI}), one-way propagation and birefringence effects are expected to occur in Born-Infeld electrodynamics. 

\begin{figure}[t]
    \includegraphics[width=\linewidth]{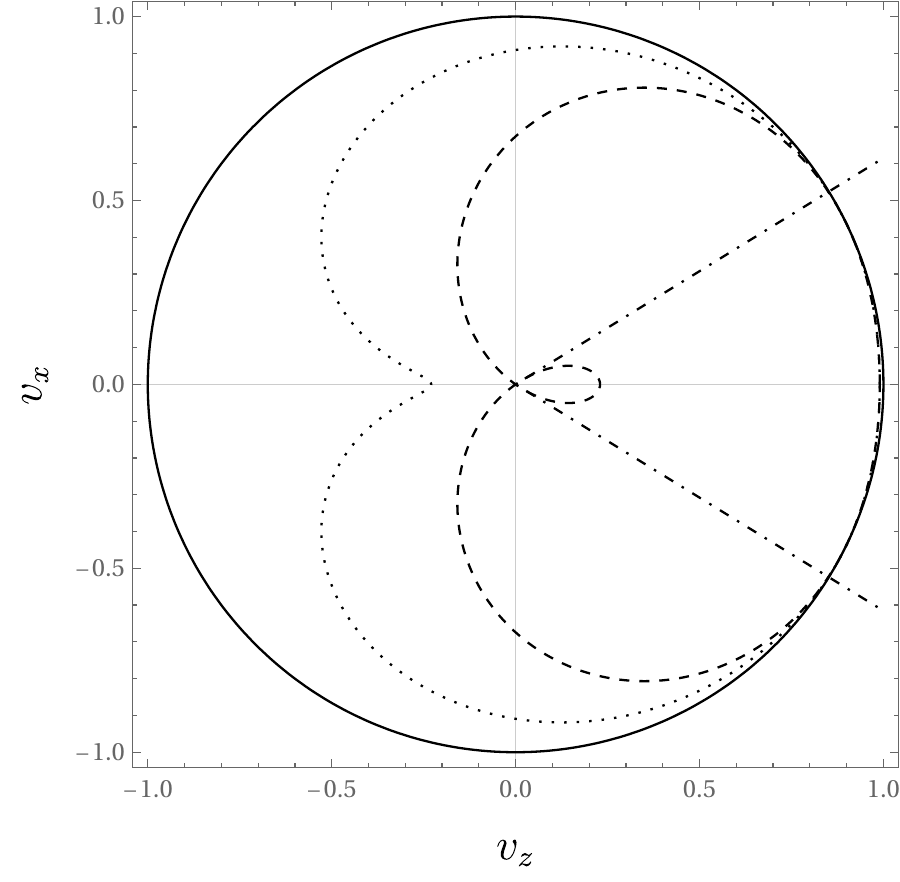}
    \caption{Normal surfaces for wave propagation in plane $xz$ in Born-Infeld theory. Here we set $E = 1.22 \beta$ and $B =1.09 \beta$. Note that, within the angle defined by the two dot-dashed lines, the phase velocity will present two solutions in a given direction, thus exhibiting birefringence. Additionally, in this sector, there is no solution in the opposite direction, which is a feature of one-way propagation. The dotted curve represents the magnitude of the group velocity, and it was added for the sake of comparison.}
    \label{figBIoneway}
\end{figure}
The normal surface for phase velocity is represented by the dashed curve in Fig.~\ref{figBIoneway}. The angle within the two dot-dashed lines defines a sector for which both one-way propagation and the birefringence occur. Notice that there is no solution in the opposite direction in this region.  Once the distance from the origin to a given point in the dashed curve indicates the magnitude of the phase velocity, it can be seen that for certain directions of propagation, the phase velocity is closer to zero, indicating that slow-light solutions are possible. 

The angles at which the phase velocities are zero are given by
\begin{equation}
  \cos ^{-1}\left(\pm \frac{\sqrt{E^2-B^2 +\mathcal{J}_0}}{E}\right).
    \nonumber
\end{equation}
This angle only exists when $E \geq \beta$, as is the case illustrated in this figure. 
Furthermore, phase and group velocities are zero for both $E=\beta$ and $\theta=0$. 
The dotted curve in this figure represents the magnitude of the group velocity, and it was added for the sake of comparison.

\subsection{ModMax Theory}
A recently proposed nonlinear modification of Maxwell electrodynamics \cite{2020PhRvD.102l1703B}, known as ModMax, is described by the Lagrangian density 
\begin{align}
L_{MM}=-\frac{\cosh\gamma}{4} F  +\frac{\sinh\gamma}{4} \sqrt{F^2+G^2},
\end{align}
where $\gamma$ is a parameter that must satisfy $\gamma \ge 0$ to ensure unitarity and causality to the theory. This parameter measures the strength of the nonlinearities of the theory and may introduce significant modifications to Maxwell's electrodynamics in a regime of strong fields. For example, solutions for wave propagation will depend on $\gamma$, which is one of the possible ways to find experimental support for this proposal. Naturally, we must assume that $\gamma$ is sufficiently small in order to make the nonlinear theory compatible with classical electromagnetism in the regime of weak fields. 

In the domain of such nonlinear electrodynamics, the possible propagating modes are determined by birefringence coefficients $\xi_{\scriptscriptstyle +}=0$, which leads to an ordinary mode that propagates with speed $c$, and 
\begin{align}
\xi_{\scriptscriptstyle -} = -\frac{4\tanh\gamma}{\tanh\gamma\, F + \sqrt{F^2+G^2}},
\end{align}
which leads to an extraordinary mode whose phase velocity is given by
\begin{align}
    \begin{array}{l}
    v_{\scriptscriptstyle \pm}=\frac{1}{E^2 \xi_{\scriptscriptstyle -}-1} \Bigg[-B E \xi_{\scriptscriptstyle -} \cos \theta \pm \\[1ex]
    \sqrt{B^2 \xi_{\scriptscriptstyle -} \cos^2 \theta-(E^2\xi_{\scriptscriptstyle -}-1)\left[1+(B^2-E^2)\xi_{\scriptscriptstyle -}  \sin^2\theta\right]} \Bigg].
    \end{array}
\end{align}

If we use Schwinger critical fields as the limit of validity of Maxwell's electrodynamics, a rough estimate of the maximum value of $\gamma$ can be obtained. For simplicity, let us consider the particular case where $G=0$, and compare the QED expansion for $\mu F \ll 1$, given by $ L_{QED}=-\frac{F}{4}(1-\mu F)$, 
with the ModMax expansion ($\gamma \ll 1$), given by $L_{MM}=-\frac{F}{4}(1-\gamma)$.
Thus, in such regime, the ModMax parameter $\gamma$ should be compared with the correction brought about by QED, which leads to $\gamma \lesssim 2.6\times10^{-5}$. Larger values of $\gamma$ would lead to significant modifications of classical electrodynamics, which we would like to avoid in this study. In a more rigorous procedure, $\gamma$ can be constrained by means of experiments that involve precision measurements, e.g., in the study of light propagation in a scenario of strong fields.

\begin{figure}[t]
    \includegraphics[width=\linewidth]{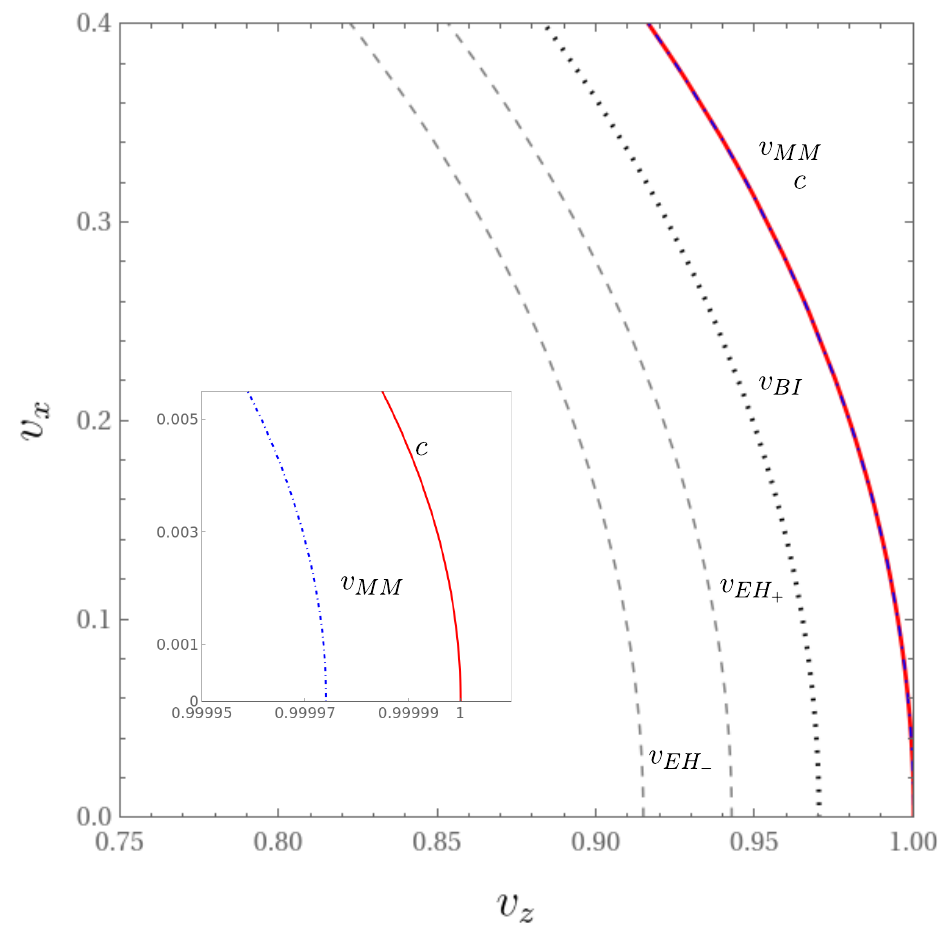}
    \caption{Normal surfaces for wave propagation in $xz$ plane. The field strengths were chosen to be $B=10^{11}$T and $E=10^{13}$Vm$^{-1}$ (in SI units), which are values typically appearing in the description of certain magnetars \cite{2022ApJ...926..111W,2013Natur.500..312T,2021Univ....7..351I}. The two dashed curves correspond to the EH phase velocities ($v_{EH_{\scriptscriptstyle \pm}}$), the dotted curve represents the Born-Infeld phase velocity, and the dot-dashed curve corresponds to the ModMax phase velocity. Here, we used  $\gamma=2.6\times 10^{-5}$. The inset provides a magnified view, comparing the ModMax correction to the classical linear theory.}
    \label{figMM}
\end{figure}

To examine the behavior of the phase velocity, let us assume the same field configuration chosen in Sec. \ref{EHLsection}: crossed fields, with an electric field aligned in the $x$ direction and a magnetic field in $y$ direction. Furthermore, expressing the unit directional vector $\hat n$ in spherical coordinates and choosing the propagation in the $xz$ plane, we obtain the behavior depicted in Fig.\ref{figMM}, where we compare the phase velocities for the three theories in consideration. It can be inferred from this figure that the kind of correction provided by ModMax electrodynamics is less sensitive to the presence of strong fields than it occurs in the other nonlinear theories considered here. This is because in the ModMax Lagrangian, the linear limit is not obtained by an expansion in the fields, as occurs with the other nonlinear models. Then, the order of the invariants in such a model will always be the same, no matter how many terms you keep in the expansion for $\gamma \ll 1$. Notice that, for the field strengths considered in the figure, the Euler-Heisenberg model gives the largest correction to the Maxwell electrodynamics, followed by Born-Infeld. ModMax correction becomes apparent only in a magnified plot, as illustrated in the inset figure. 

\section{Final remarks}
\label{final}
Here, several aspects of light propagation in nonlinear theories for electromagnetism formulated in terms of the two Lorentz invariants, $F$ and $G$, are investigated. Expressions for phase and group velocities were derived for general nonlinear theories, and certain regimes were analyzed. In particular, it is shown that when the birefringence parameter $\xi_{\scriptscriptstyle \pm}$ is treated as a small perturbation, the magnitudes of the phase and group velocities will coincide, independent of the specific nonlinear theory considered. As a consequence, in such a regime, a wave packet propagates at the same speed as plane waves. However, their directions of propagation generally do not align. 
Furthermore, when both electric and magnetic external fields are present, birefringence and non-reciprocal phenomena may occur. Birefringence has already been well studied in the literature and may formally occur whenever two distinct values for $\xi_{\scriptscriptstyle \pm}$ are allowed. However, as discussed in Sec.~\ref{sec-BI}, even when only one value $\xi_{\scriptscriptstyle \pm}$ is allowed, birefringence may arise as a spin-off of one-way propagation phenomenon. Regarding non-reciprocity, even under strong electromagnetic fields, such as $E=10^{13}$Vm$^{-1}$ and $B=10^{11}T$, this phenomenon appears as a very weak effect, yielding corrections of the order of $10^{-7}-10^{-8}$ for Born Infeld, Euler-Heisenberg, and ModMax nonlinear theories. 


An interesting aspect of Born-Infeld theory is the existence of a fundamental limitation on the electromagnetic field strengths. For example, in the absence of a magnetic field, it is straightforward to conclude from the Lagrangian density that the maximum value of the electric field strength is $\beta$. This result coincides with the maximum value of the electric field produced by a point charge.  However, when both fields are present, the usual bound does not necessarily apply when $\mathbf{E}$ and $\mathbf{B}$ are not parallel, as an inspection of Eq.~(\ref{conditionBI}) reveals. Thus, fields exceeding $\beta$ seem to be possible in this nonlinear electrodynamics \cite{PhysRevD.87.087703}, as discussed in some detail in Sec.~\ref{sec-BI}, and this fact is relevant for applications involving extreme electromagnetic environments, such as those found in magnetars. An important consequence of allowing electric and magnetic fields to go beyond $\beta$ is the appearance of slow light, one-way propagation, and birefringence phenomena. This possibility was already anticipated in the context of artificial optical materials \cite{PhysRevA.89.043822}.

\acknowledgments
T.~W.~C. acknowledges CAPES for the scholarship provided during his MSc. studies. V.~A.~D.~L. is supported in part by the Brazilian research agency CNPq under Grant No. 302492/2022-4. E. G. H. acknowledges CNPq under Grant No. 151974/2024-1. C.C.H.R. would like to thank the FAPDF ({\it Funda\c{c}\~ao de Apoio \`a Pesquisa do Distrito
Federal}) under Grant No. 00193-00002051/2023-14.


\bibliography{ref}

\end{document}